# Service-based Routing at the Edge

Dirk Trossen
InterDigital Europe, Ltd
London, UK
dirk.trossen@interdigital.com

Sebastian Robitzsch
InterDigital Europe, Ltd
London, UK
sebastian.robitzsch@interdigital.com

Scott Hergenhan
InterDigital Inc.
Conshohocken, US
scott.hergenhan@interdigital.com

Janne Riihijärvi
Institute for Networked Systems
RWTH Aachen University, Germany
jar@inets.rwth-aachen.de

Martin Reed
University of Essex
Colchester, UK
mjreed@essex.ac.uk

Mays Al-Naday
University of Essex
Colchester, UK
mfhaln@essex.ac.uk

## ABSTRACT

Future scenarios, such as AR/VR, pose challenging latency and bandwidth requirements in 5G. This need is complemented by the adoption of cloud principles for providing services, particularly for virtualizing service components with which virtualized instances can appear rapidly at different execution points in the network. While providing service endpoints close to the end user appears straightforward, this early service break-out is currently limited to routing requests to Point-of-Presence (POP) nodes provided by a few global CDN players deep in the customer network. In this paper, we propose instead to turn the edge of the Internet into a rich service-based routing infrastructure with services being provided through edge compute nodes, without needing indirect routing. Our approach interprets every IP-based service as a named service over a (L2 or similar) transport network, requiring no per-flow state in the network, while natively supporting both unicast and multicast delivery. The solution allows route adjustments in time scales of few tens of milliseconds, enabling rapid failure recovery, extremely responsive load balancing, efficient mobility support, and more. We implemented our solution on standard SDN-based infrastructure and in mobile terminals in a backwards-compatible manner, enabling a performance evaluation that shows significant improvements in network utilization as well as flow setup times.

## 1 Introduction

The emergence of 5G systems facilitates many use cases, particularly those that rely on low latency communication as well as high bandwidth for delivery of, e.g., virtual reality content. Service execution in 5G is assumed to be highly flexible, driven by the adoption of cloud design principles which see services being provided potentially near end users. Examples for such use cases are 5G mobile networks with service-based control plane architectures, those control planes being realized over a pure software-defined Layer 2 network that interconnects regional data centers. In addition, services for virtual reality, industrial as well as vehicular applications at the very edge of the network are being pursued in work on Multi-access Edge Computing (MEC) or Fog computing; these are driven by the edge densification in terms of network and computing capabilities that is being outlined not only for 5G but also beyond.

Catering for these stringent requirements of low latency, high throughput and flexibility in service execution has become increasingly challenging since early service termination is limited to providing requests to well-managed points-of-presence (POPs) deep in the customer network, while providing services closer to end users would need to rely on service routing capabilities at the edge where IP routing has not been established yet. Furthermore, traffic in the present-day TCP/IP stack design needs to traverse several layers in a number of elements for the provisioning of an end-to-end service, while rigid service bindings make it difficult to quickly redirect relations to nearer service execution endpoints. *This highly motivates solutions that turn the current Layer 2 access network into an environment that can flexibly route (Internet) service requests.*

We propose to address these challenges by radically flattening the Internet protocol stack with each Internet service residing directly on top of a Name-based Routing (NbR) layer. Such direct mapping of HTTP and other services, including IP itself, will allow for utilizing Layer 2 multicast forwarding capabilities, significantly reducing bandwidth requirements for multi-user HTTP-based services such as those employed for virtual reality to name just one example. At the infrastructure level, we utilize the recent introduction of standardized programmable forwarding solutions, such as OpenFlow-based SDN switches. These technologies make changes in the network infrastructure a matter of updating the controlling software, as opposed to expensive hardware replacements. We use this foundation to realize an efficient *path-based forwarding* mechanism within an operator network, enabling extremely responsive and efficient name-based routing of Internet services within an autonomous domain. Importantly, the solution retains full compatibility with the existing Internet protocol interfaces towards peering networks and legacy devices. We show this can be accomplished without per-flow state in the network, and with computational requirements similar to existing networks. The solutions outlined in this paper has already been recognized as a possible deployment choice for Release 16 of the 3GPP 5G control plane [14], while newer work pursues the adoption for user plane services.

The rest of this paper is structured as follows. In the following section, we discuss the opportunities enabled by our approach. We then present the system architecture as well as design details in Section 3. In Section 4 we provide selected evaluation results to compare against the opportunities outlined in Section 2. Finally, we discuss experiences from several trial deployments based on our prototype in Section 5, before concluding in Section 7 after a brief related work overview in Section 6.

## 2 Opportunities to Test Against

In this section, we briefly discuss some of the key enabling opportunities provided by service routing directly on L2. How these opportunities are then addressed by our proposed system is discussed in Section 3, while we report in Section 4 evaluation results from testbed deployments as well as dedicated simulations to further quantify the gains these opportunities can achieve.

### 2.1 Multicast Delivery of HTTP Responses

The vast majority of current Internet traffic is due to unicast delivery of relatively immutable content such as video or software to very large client groups [9]. This has resulted in significant redundancy in network traffic, possibly creating capacity bottlenecks both in the core network as well as the server infrastructure serving the content. Technologies such as content delivery networks (CDNs) help to spread out the network load, but are complex to manage, have inherent limits in terms of how rapidly they can react to changing network and server conditions, and cannot fundamentally reduce the network overhead arising from redundant unicast streams. Furthermore, CDNs traditionally only reach into POPs within customer networks, therefore they do no reduce the load of transfer from said POP to the end customers in that edge network.

Our solution, in contrast, enables *opportunistic multicast delivery* of content, automatically delivering responses to quasi-concurrent requests in a single lightweight multicast transmission over the L2 customer network, extending the reach of CDN like services closer to the end user. Unlike traditional IP multicast, our approach has no additional setup time overhead and it does not require per-flow state in the network. The time period (which we call the *catchment interval*) over which this process takes place can be flexibly configured on a per-service basis, further improving the opportunity for multicast delivery. For latency-sensitive services, such as video chunk delivery, short catchment intervals are appropriate (100 – 1000 ms for example), whereas for delivering software updates, carrying out database or cloud service synchronization, and other relatively delay-tolerant services, much longer catchment intervals can be used. The gains from multicast delivery can be especially dramatic for highly popular content at peak request times, e.g., a new episode of a popular series becoming available. As an optimization, that can again be enabled on per-service basis, we can combine opportunistic multicast delivery with *request suppression* where the origin server does not even receive the requests for content that are opportunistically multicasted, thereby reducing the server load and costs for content delivery even further.

### 2.2 Flow Setup

One of the key latency bottlenecks in the current Internet is caused by the high flow setup latency, especially when transport (or higher) layer security is involved. Furthermore, many applications still rely (for reliability reasons and to simplify development) on non-persistent connections that get rebuilt for every individual request for each content item, even when served by the same origin server. In contrast, our proposal enables (but does not require) splitting of the connection at the network ingress point. Since this is usually very close to the end users, in terms of latency, optimizing the residual latency in the core translates to substantial latency reduction at the edge, even if the client-to-edge connection establishment is not modified. Such approaches have been successfully used in the wireless community to deal with extreme latencies (as found in satellite communications for example [24]), and our approach enables deploying them transparently at the network edge.

### 2.3 Service Indirection

Another key latency in the Internet is that for service resolution, performed through the Domain Name Service (DNS). Services accessed via their URL, say at foo.com, are translated to an IP address as a routing locator. The *initial lookup* of sites causes aggregated latency in particular in scenarios where many such URLs are embedded into a single site that is accessed by the user, e.g., through advertisements inserted into the webpage. Through this, DNS latency can become a rather significant latency factor in the overall end user experience.

Also important, in particular in edge network scenarios, is the *latency for service indirection*, i.e., directing traffic to an already resolved name but with a different location in the network. While content delivery networks (CDNs) provide approaches to indirection through appropriate DNS record configurations, those relations are static in terms of relating the indirection to the specific CDN site at which the service is ultimately provided, while data center internal load balancing mechanisms are used to dispatch the incoming request to the appropriate (internal) computing resource. An edge network, however, constitutes a distributed approach to data centers, where compute resources can be located at smaller, yet distributed sites.

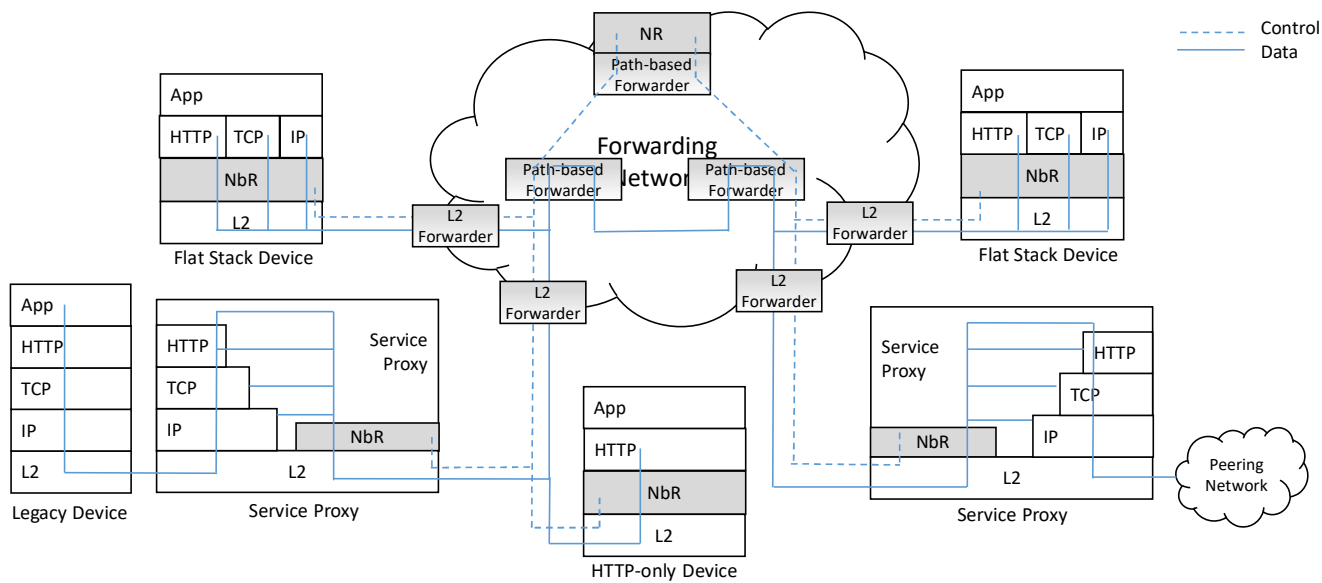

**Fig.1: Edge Network Architecture**

Through orchestration of services, computing instances can be placed into those distributed sites relatively quickly (in order of a few seconds while pre-placed instances can be 'activated' even faster). Relying on the DNS poses a problem here since reconfiguring the relevant DNS entries is too slow to support the fast provisioning times of virtualized service instances. Furthermore, caching DNS entries at clients is not standardized and can therefore lead to stale entries of mapping from service names to routing locators of previous service instances. A detailed analysis of the impact of DNS on service mapping has been provided in [25].

Our solution will address this issue be providing a reactive service registration, discovery and update protocol that integrates with the DNS for those services not provided in the (edge) network, i.e., the Internet. With this, we aim at achieving initial lookup times that improves against DNS resolvers in the Internet while achieving service indirection to new service instances at even faster timescales.

## 3 System Description

In the following, we provide more details on the key aspects of our network architecture, the operations in our name-based routing and the forwarding of packets in an end-to-end manner over a Layer 2 network. We further provide details on the end-to-end flow management as well as the integration with the transport network infrastructure.

### 3.1 Edge Network Architecture

Key to our approach is that Internet services are being interpreted as the main unit of transfer in our architecture shown in Figure 1. For this, we treat any Internet service as a sequence of *named service transactions* (NST) which is in turn suitably routed over the *NbR layer*. As a result of this name-based interpretation, the protocol stack in end devices flattens to four layers with Internet services and NbR building on top of layers 1 and 2; we call these devices *flat stack* devices, realized over the proposed NbR stack. These *flat stack* devices enable the possibility (but not requirement) of pure application protocol devices, e.g., HTTP- or CoAP-only devices, not relying on the existence of the TCP/IP stack in the device.

However, we also preserve the interfaces to legacy TCP/IP stack devices and peering networks through *service proxy* devices. These *service proxy* devices terminate a traditional Internet protocol stack communication and translate it into a resulting flat protocol transaction based on the operations defined in Section 3.3. Termination here can be based on well-known port numbers for specific treatment of certain Internet protocols, ultimately falling back to the IP datagram service as the minimal service being mapped.

The end-to-end packet forwarding is described in Section 3.2, outlining the operations of the Layer 2 and path-based forwarders. The forwarders utilize topology knowledge obtained by a Name Resolver (NR) during the initial bootstrapping of the transport network. Note that the NR referred to here is that within the name-based routing mechanism described in the 3GPP 5G Release 16 (Annex G.4) [14].

The details of the operations of the NbR layer are presented in Section 3.3. As explained in that section, the NbR uses an interaction with the NR, extending the interaction with the DNS in IP routing to a reactive name registration and discovery mechanism for any IP-based protocol.

An important aspect of our architecture is the mapping of the end-to-end flow semantic, established in many Internet services, onto the flat protocol stack. Section 3.4 outlines our flow management that exists between the end devices.

## 3.2 End-to-End Packet Forwarding

Packets in our edge network architecture of Figure 1 are forwarded at Layer 2 in an end-to-end manner between islands of local LAN connectivity, using a hybrid forwarding solution outlined in the following.

For our solution, we adopt the view of a so-called *vertical LAN*, being currently developed in 3GPP [14] for the next release of 5G. In this view, a set of services is configured to be provided through a specific instance of a vertical LAN, utilizing a LAN identifier at the lower level of forwarding, similar to a virtual LAN tag in existing fixed networks. With this, each vertical LAN represents an intranet with Internet connectivity being established through a dedicated packet gateway (see Figure 1). Local wireless and fixed connectivity to the network ingress (L2 forwarder) is realized through an Ethernet-type abstraction with link-level specific solutions being utilized, e.g., for cellular, WiFi or fixed networks. Through this, fixed, wireless, and cellular connectivity is converged into a single Layer 2 abstraction with MAC identifiers being used as end-to-end identifiers, while realizing the distribution across local connectivity islands via Layer 2 forwarders (see Figure 1). With this, L2 forwarders only hold MAC information for locally attached (L2) devices, avoiding the scalability problems of routing on flat labels [15] across ALL possible devices in the network.

For the communication between two L2 forwarders, *path-based forwarding* is used. Here, the path between two L2 forwarders is encoded through a *bitfield*, provided in the packet header as explained later. Each bit-position in this bitfield represents a unique link in the network. Each such bit-position is assigned by the NR during the bootstrapping of the network. Upon receiving an incoming packet, each path-based forwarder in Figure 1 inspects the bitfield for the presence of any local link that is part of the path and connected to an output port. The presence check is implemented via a simple binary comparison operation. If no link is found, the packet is dropped. Such bitfield-based path representation also allows for creating multicast relations in an ad-hoc manner by combining two or more path identifiers through a binary OR operation, as explained later in our example HTTP mapping onto the NbR layer. Note that due to the assignment of a bit-position to a link, path identifiers are bidirectional and can therefore be used for request/response communication without incurring any need for path computation on the return path.

We now show how end-to-end forwarding is enabled by combining the local LAN forwarding with the path-based forwarding between LAN islands. Consider a packet that is sent from one Layer 2 device connected to one L2 forwarder then through the path-based forwarding before finally reaching a device connected to another L2 device. To fully determine the path, the sending device provides the MAC address of the end destination while also providing the path identifier to the ingress L2 forwarder. To determine the latter, the interaction with the NR is used, as required by the protocol-specific mapping of the named service transaction to the end-to-end forwarding (see Section 3.3 for the HTTP example of our mapping). The ingress L2 forwarder then performs the path-based forwarding towards the egress L2 forwarder with the intermediary path-based forwarder performing the necessary bit position checks to forward the packet. Upon arrival at the egress L2 forwarder, the destination MAC address is used for the link-local transfer. Note that certain LAN technologies, such as WiFi, will require the use of the ingress/egress MAC addresses for intermediary (island) LAN communication, leading to a destination MAC re-writing at the ingress/egress.

For this end-to-end transfer, the general packet structure of Figure 2 is used. The *Name_ID* field is used for the NbR operations, explained in Section 3.3, while the payload contains the information related to the transaction-based flow management described in Section 3.4.

| Src MAC | Dst MAC | pathID | NAME_ID | Payload |
|---------|---------|--------|---------|---------|

**Fig.2: General Packet Structure**

An emerging technology for Layer 2 forwarding that suits our architecture is that of software-defined networking (SDN) [1], which allows for programmatically forwarding packets at Layer 2. Switch-based rules are executed with such rules being populated by the SDN controller. Rules can act upon so-called *matching fields*, as defined by the OpenFlow protocol specification [2]. Those fields include Ethernet MAC addresses, IPv4/6 source and destination addresses and other well-known Layer 3 and even 4 transport fields.

As shown in [3], efficient path-based forwarding can be realized in SDN networks by placing the aforementioned bitfield based path identifiers (the PathID in Fig 2) into a newly added IPv6 source and destination field of a forwarded packet. Utilizing the IPv6 source/destination fields allows for natively supporting 256 links in a transport network. Larger topologies can be supported by extension schemes but are left out of this paper for brevity of the presentation. As mentioned before, the NR assigns to each link at each switch a unique bit-number in the bitfield during network bootstrapping. In order to forward based on such bitfield path information, the NR instructs the SDN controller to insert a suitable wildcard matching rule into the SDN switch. This wildcard at a given switch is defined by the bit-number that has been assigned to a particular link at that switch during bootstrapping. Wildcard matching as a generalization of longest prefix matching is natively supported by SDN-based switches since the OpenFlow v1.3 specification and efficiently implemented through hardware TCAM based operations. This switching mechanism means that the number of SDN forwarding rules at each switch only depends on the number of output ports at each switch; it also means that it can transport any number of higher-layer flows over the same transport network without specific flow rules being necessary. This results in a constant forwarding table size while no controller-switch interaction is necessary for any flow setup; only changes in forwarding topology (resulting in a change of port to bit number assignment) will require suitable changes of forwarding rules in switches.

Although we focus our presentation on Layer 2 forwarding approaches compatible with our architecture, path-based transport networks can also be established as an overlay over other Layer 2 network solutions. For instance, the BIER (Bit Indexed Explicit Replication) efforts within the Internet Engineering Task Force (IETF) establish such path-based forwarding transport as an overlay over existing, e.g., MPLS, networks [4]. The path-based forwarding identification is similar to the aforementioned SDN realization although the bitfield represents ingress/egress information rather than links along the path.

Yet another transport network example is presented in [5], utilizing flow aggregation over SDN networks. The flow aggregation again results in a path representation that is independent from the specific flows traversing the network.

## 3.3 Operations of the NbR Layer

In this section, we will outline the general operations of the name-based routing layer before illustrating the operations with the example of HTTP-based services.

### 3.3.1 General Operations

We define the semantics of our name-based routing as that of a publish-subscribe system over a name. In our realization, we use *structured names* in a tree structure with the root specific to the (Internet) service name, such as a URL, and can therefore derive the matching semantics directly from the name.

The intention to receive packets with a certain name is expressed through a subscription while sending packets to a name is expressed through a publication. The matching of a sender to a receiver is realized through the name resolver (NR) in Figure 1. The exact nature of the matching is defined through the semantics of the service and, therefore, through the nature of the name provided. For instance, HTTP and raw IP services are matched to exactly one subscriber only, providing an anycast capability, while IP multicast services are matched against any subscriber (with the IP multicast address being the name).

### 3.3.2 API to Upper Layers

The pub/sub operations of the NbR layer are exposed through the following API calls:

```
conn = send(name, payload)

send(conn, payload)

conn = receive(name, &payload)

receive(conn, &payload)
```

The first send() call is used for initiating a send operation to a name with a connection handle returned, while the second send() is used for return calls, using a connection parameter that is being received with the receive() call to an incoming connection or for subsequent outgoing calls after an initial request to a name has been made. A return send() is received at the other (client) side through the second receive() call where the conn parameter is obtained by the corresponding send() call for the outgoing call. With these API functions, we provide means for providing name-based transactions with return responses association provided natively.

In our realization, the conn parameter represents the bitfield used for path-based forwarding in the remote host case or the hash of the local MAC address in case of link-local connections (with bitfield of remote forwarding set to 0).

### 3.3.3 Registration and Discovery of IP-based Services

IP-based services are exposed through the subscription to a name specific to the service, which in turn is realized through a *registration protocol* between an end device hosting the service and the NR shown in Figure 1. Upon registration, the NR stores reachability information that is suitable for path calculation between the exposing and the requesting device. The path calculation can be over shortest paths, used by default in our implementation. However, the path-based mechanism used here opens up to other possibilities including using highly optimal traffic engineering solutions as shown in [24]. In our realization, we use network domain unique *host identifiers* that are assigned to end devices during the connectivity setup together with the MAC address of the hosting end device.

Sending a packet of a given IP-based service is realized through a *discovery protocol*, which returns suitable reachability information, previously stored as part of the aforementioned registration process [1]. This reachability information consists of a suitable `pathID`, i.e., the forwarding information between ingress and egress L2 forwarder at which both the hosting and the request devices are connection. Furthermore, the destination MAC address of the hosting end device is returned. This combined reachability information can then be used in the general packet structure of Figure 2 to forward the packet to the destination, as explained in Section 3.2. To reduce latency in further communication between two devices, the forwarding information is locally cached at the end device. This cached forwarding information is maintained through *path updates* sent by the NR to avoid stale forwarding information. When a hosting end device moves or de-registers, path updates are triggered to be sent, therefore implementing our solution for the service indirection opportunity presented in Section 2.3.

### 3.3.4 Mapping HTTP onto NbR

In the cases of devices that wish to act as *flat stack devices,* the Internet service layers, such as the HTTP protocol stack or the TCP protocol stack, are adapted to run on top of this new API, implementing the semantics of the respective Internet protocol through suitable transactions at the name level. In the example of HTTP, the standard operations of DNS resolution for the server to be contacted and opening of a TCP socket are altogether replaced by a single `send(FQDN, HTTP request)` call, while the response will be sent by the server, which received the request through a `receive(FQDN, &payload)` call, using the returned `conn` parameter to send the response with the second `send()` API call[2].

### 3.3.5 Name-Based Ad-Hoc Multicast for HTTP

Section 2.1 pointed out the opportunity to deliver HTTP responses in efficient multicast. We realize this opportunity by sending the same payload (i.e., our HTTP response to the same resource across a number of pending requests) to multiple clients using the inherent ad-hoc multicast made possible through the path-based forwarding mechanism described in Section 3.2. This is achieved simply by performing a binary OR combination of the `conn` parameters received in the incoming requests via the `receive()` function[3]. What is required in the HTTP stack implementation (of the flat stack device or service proxy) is a logic to decide that if two or more outstanding requests are possible to be served by one response. For this, upon receiving an incoming request, the HTTP stack determines any outstanding request to the same resource. 'Same' here is defined as URI-specific combination of the request URI and URI-specific header fields, such as browsing agent or similar, called `requestID` in the following. For the sake of brevity, we leave out the distribution of those URL-specific rules to the originating and serving endpoints.

Once a determination is made that two, or more, requests are requesting the same resource, i.e., are having the same request ID, the HTTP stack maintains a temporary mapping of the request ID to the respective `conn` parameters delivered by the `receive()` call. Upon receiving the HTTP response from its application-level logic, the HTTP stack will generate the suitable `send(conn, payload)` call where the provided `conn` parameter (the pathID) is a bitwise OR of all previously stored conn parameters received in the `receive()` call. The NbR layer will recognize the use of those ad-hoc created `conn` parameters and set the destination MAC address in the general packet structure of Figure 2 to the Ethernet broadcast MAC address[4] as the destination address. This leads to sending the response to all end devices at the egress L2 forwarders to which the

---

response will be forwarded based on the combined `conn` parameter.

For the local end devices at those egress L2 forwarders to determine the relevance of the response received at the broadcast channel, the HTTP stack of the serving endpoint includes the aforementioned `requestID` into the payload of the packet (see Figure 2), while the originating endpoint maintains an internal table with the `requestID` of pending requests and its associated `conn` handle[5]. If no matching `requestID` is found, the packet is not delivered to the NbR layer of the incoming device. If a request is found, the NbR layer delivers the response via the `receive()` call, using the `conn` handle stored in the pending request table.

### 3.4 Flow Management

Mapping flows onto Layer 2 transport relations is an essential function; this can be carried out either in the service proxy device or directly in a flat stack device. For this, we utilize the path-based forwarding characteristic of our transport network to split the flow/congestion management from a pure end-to-end mechanism to one that manages resources end-to-NbR (i.e., from device to proxy device or application to device-internal NbR layer and vice versa) as well as NbR-to-NbR layer, as outlined in Figure 2. With this, anything transported via the NbR layer from one device to another is managed by the same flow management regime.

Such a split in the flow/congestion management brings several advantages. Firstly, given the device or link-local nature of the TCP flow setup at both sides of the split resource management regime, flow setup is limited to either side and therefore local only. Hence, we can expect that, e.g., flapping behavior of terminating and re-established TCP connections is limited to the side where such flapping is being realized. Section 4.2 will investigate the benefits of this split flow/congestion management in more detail.

Secondly, resource fairness is preserved end-to-end, while allowing for optimizing the resource management of the intermediary Layer 2 transport network and the NbR-to-NbR interactions. Key to such an ability is the mapping of end-to-NbR flows onto NbR-to-NbR flows. In this mapping, NbR-to-NbR flow/congestion management is maintained over so-called 'managed service flows' as longer-term relationships at the NbR layer. End-to-NbR flow/congestion management observes the well-known TCP-friendly regime known in the

Internet with the result being an end-to-end TCP-friendly resource management.

Figure 3 shows the mapping we propose to be realized at the NbR layer in more detail. The mapping of end-to-NbR flows onto the longer-lived named service flows between NbR-based devices is shown as a mapping from so-called 'IP transactions' onto 'named service transactions', with the latter being managed by the flow control of the longer-lived (named) service flow. The nature of the IP transactions depends on the supported protocol mapping at the NbR-based device (with Section 3.3 outlining an HTTP-level mapping example), assuming at the very least an IP-based mapping to exist. We assume that the access to the named service flow send buffer is equally shared among all locally incoming IP transactions with the max MTU size being the granularity of the send buffer.

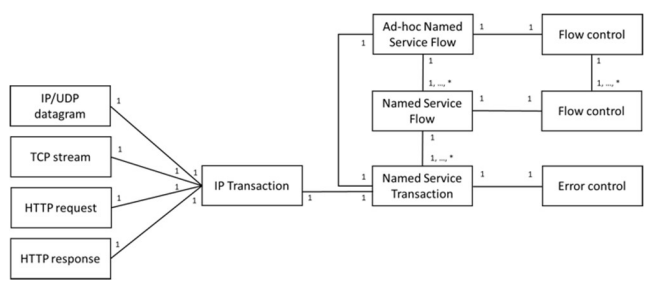

**Fig.3: IP to named service flow mappings**

### 3.5 Security Considerations

In this section, we will discuss two security aspects arising from our design, namely the handling of secure end-to-end transport connections as well as the protection of forwarding security.

### 3.5.1 Transport Layer Security

For the realization of transport layer security (TLS), we differentiate two cases, namely the provisioning between (i) two intranet devices and (ii) at least one Internet-based device. For the former case, we mimic the standardized TLS handshake for the first secure transaction towards an (intranet based) HTTP service, leading to the security credentials to encrypt the payload in Figure 2, i.e., the HTTP requests and responses exchanged between the client and the HTTP-based service.

For the case of an Internet-based service, the proxy device in Figure 1 towards the Internet may act as a TLS proxy towards the service FQDN provided in the TLS handshake. Alternatively, certificate sharing agreements can exist between the service proxy provider and the (Internet) service provider being addressed in the request, similar to existing arrangements between content and CDN providers. In that case, the service proxy can apply the necessary logic for, e.g., handling multicast responses towards the intranet-based clients having requested the same content from the Internet-based service.

---

[5] Checking such internal table can be efficiently realized at a low driver level or even be HW offloaded for performance optimization; this is further investigated in our future work.

In some cases, e.g., online banking, there may be no certificate sharing arrangement. In this case the traffic is handled opaquely by using the default IP mapping in our solution. In this case clearly there would be no multicast advantages as the traffic is inherently unicast in nature.

### 3.5.2 Forwarding Security

One key aspect to forwarding security is the avoidance of malicious sending of traffic along parts of the network not destined for delivery. In our proposed path-based forwarding, such wrongful delivery can take place when bit positions are set in the pathID bitfield that were not originally part of the pathID assigned to the specific delivery. The work in [6] outlines an approach to protect the path information through certificate information at the ingress of the network, which can be applied to our design.

Another aspect is that of erroneous sending, e.g., when bit positions are accidentally set in positions not previously defined in the pathID. For this, checksum checks over the pathID can be applied at the ingress, lowering the probability for such erroneous bit alteration.

## 4 Evaluation

The following section tests our solution against the opportunities outlined in Section 2 through evaluation insights, both based on simulations and prototype efforts in a lab-based network deployment.

### 4.1 Multicast Gain

#### 4.1.1 Evaluation Scenarios

To quantify the benefits of our solution to the network operator, we evaluate the network gain from multicast transmission, in terms of traffic reduction in the network. We represent this as a ratio of traffic using only unicast transmission compared to traffic using our multicast transmission; in both cases the same services are delivered to end users. To achieve this, we apply an analytical model using realistic topologies, obtained from the Internet Topology Zoo database [10]. Particularly, we use the AT&T MPLS topology shown in Figure 4 below. We assume a catalog of 1000 aggregate services, with a Zipf-based popularity distribution. The Zipfian exponent is set to 0.82, providing alignment with realistic distribution of instantaneous popularities of Torrent-like services [11]. We assume each Service Proxy having between 1000 and 5000 end-users; and, a session length of 900 seconds, in which end-users randomly issue requests for any of the services in the catalog. A subset of end-users who fall within a catchment interval (0.5 or 5 seconds) are deemed to be in a common multicast group.

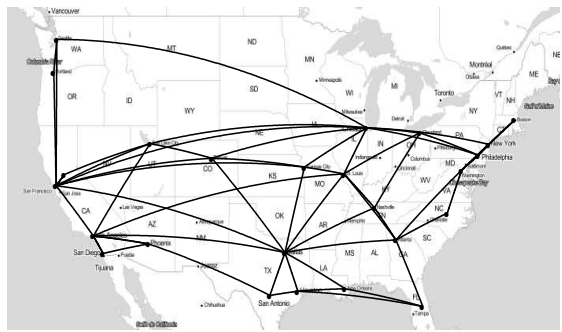

**Fig. 4: AT&T MPLS topology**

#### 4.1.2 Results

The results of Figure 5 show a considerable gain factor between ~1.6 and 2.1 for a short catchment interval of 0.5 second. This means that operators can achieve between ~40% and ~65% reduction in backhaul traffic, as the number of users per Service Proxy increases. Considering higher catchment intervals – particularly for delay tolerant applications - increases the network gain even further.

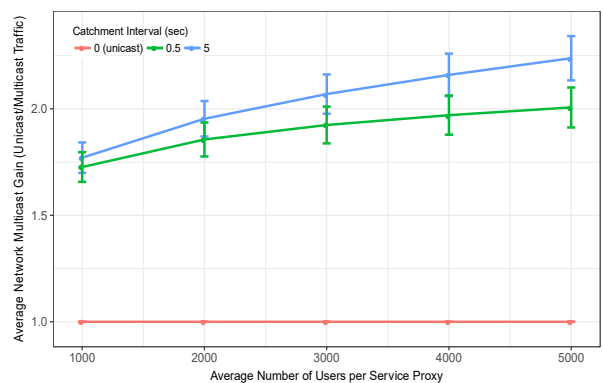

**Fig. 5: Multicast Gain**

Figure 6 shows the multicast gain achieved when increasing the Zipf exponent. This indicates higher popularity for a smaller set of items, decaying at a faster rate for each higher value of the exponent. The number of end-users attached to each service proxy is set to 3000. The results show a significant increase in multicast gain from approximately 2 to 8 with higher Zipf exponents. Notably the difference in multicast gain for different catchment intervals increases sharply with the increase in the exponent. For a catchment interval of 0.5 sec, the multicast gain grows from ~2 to 3.5; whereas for a catchment interval of 5 sec, the multicast gain grows faster than linear.

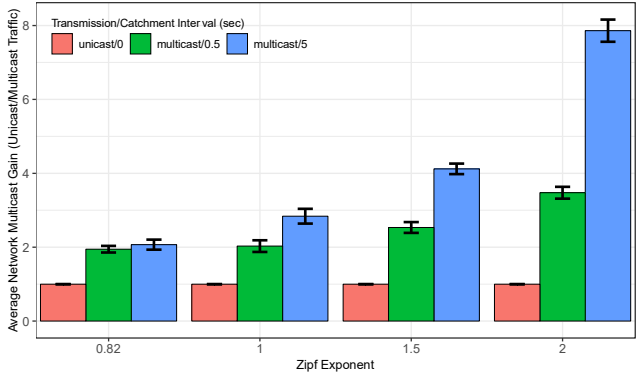

**Fig. 6: Multicast Gain for different Zipf exponent. Number of users per Service Proxy is 3000**

For transmission using unicast there is no catchment interval and no gain (i.e. a value of unity) and this is indicated in Figure 6 for comparison.

## 4.2 Flow Setup

We shall now present evaluation results for the second opportunity discussed in Section 2, namely the reduction in latency and increase in robustness for the flow setup.

### 4.2.1 Evaluation Scenarios

The MSC diagram shown in Figure 7 illustrates the differences between the flow setup behavior of classical TCP applications and the NbR approach with different deployment options.

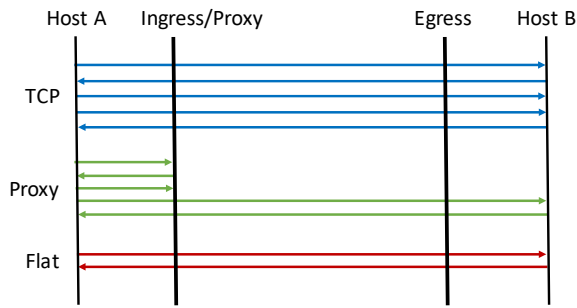

**Fig. 7: Message Sequence Charts for Flow Setup**

In the case of TCP, the flow setup messages (the three-way handshake) need to traverse the entire network, taking approximately two entire round-trip-times before the first part of the content requested from Host B reaches the requesting Host A. Using a proxy deployment allows terminating the TCP connection much closer to the originating host, requiring only the final request and response to traverse the entire network. Finally, a native deployment at the host only presents the normal socket interface towards the application, and the TCP flow setup is handled internally within the network stack of Host A, further reducing latency especially in deployments with a first wireless hop, where network latency can be appreciable.

### 4.2.2 Results

We evaluate the time from initiation of flow setup until Host A receives the first byte of content for the same topology as used in Section 4.1 (see Figure 7). We assume Host A and Host B to be located behind randomly selected ingress and egress nodes, and initially assume an average one-way propagation delay of 2ms for the ingress and egress links, and 12ms for the wide area links, with jitter set to 10% of these values (see [16] for further discussion of typical parameter settings for protocol evaluation).

The corresponding flow setup times for the three different deployment options are shown in Figure 8. Clearly a significant reduction in flow setup time compared to TCP is possible, in this case amounting to roughly halving the initial setup time (as expected from the connection setup MSC above). While the proxy solution does slightly increase latency compared to native deployment, the difference is very minor due to the small latency in the access link.

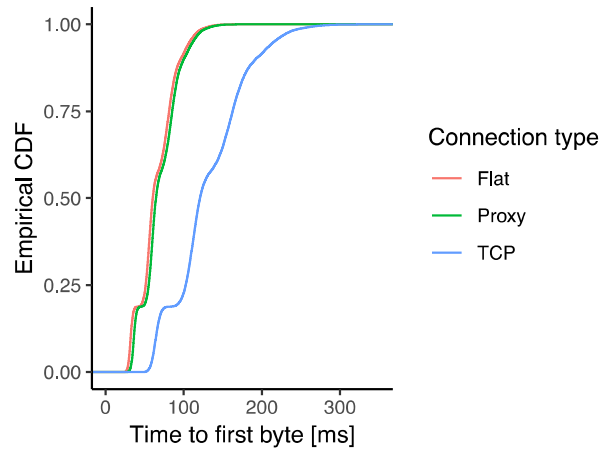

**Fig. 8: Flow Setup Times with Fast Access Links**

The role of native deployment becomes more important if the access link is wireless, as typical wireless technologies have significantly higher access delays compared to fixed networks. Figure 9 below shows the corresponding latency results assuming Host A is connected to a link with one-way propagation delay of 9 ms, value typical of modern 4G cellular networks. The difference between native and proxy-based deployment becomes much clearer in this case.

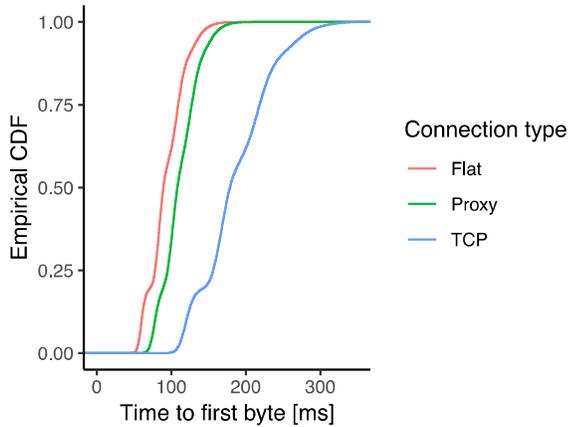

**Fig. 9: Flow Setup Times with Wireless Access Links**

Of course, the improvements in flow setup times can become much higher in absolute terms if the core network has links with very high latency. In this context our approach is very similar to the concept of *TCP proxies* that have been proposed to improve TCP performance in, for example, satellite networks. Reduction in the number of signaling messages per network hop can also bring further performance benefits through reduction of packet errors and decreasing the probability of highly delayed signaling packets. We refer the reader to [13] for a related discussion.

## 4.3 Service Indirection

We now present preliminary evaluation results for the final opportunity discussed in Section 2, namely the reduction in latency for service lookup and service indirection, based on an implementation of our network architecture (see also Section 5 for deployment insights for this platform).

In our setup, we established a core network of SDN 1Gbit/s switches, providing suitable capacity similar to that found in trial deployments we are currently conducting (see Section 5). Services were registered as `fooN.com` for N=1,…, n. We used the service proxy mode as a worst-case scenario to showcase the improvement against existing DNS-based services in terminals. In reference to Figure 10, we established a service cluster instead of the peering network at the right-hand side, while issuing service requests from the legacy device (a laptop) on the left-hand side towards service instances hosted in the cluster.

Given the network-local nature of the registration and discovery process, latency was kept to a minimum with typically less than 10ms for initial discoveries. This overall latency is comprised of about 9ms for the network latency from the service proxy (which acts as the NbR device here) to the NR at the top of the network in Figure 1 and back, while 1ms is spent in the NR on discovery of the service name and path computation from the ingress to the egress service proxy. In flat stack device operations, we expect this number to be in a similar order since even though the 9ms spent on network traversal would be reduced due to removing the need for the service proxy, more hops might be added albeit at the fast Layer 2 forwarding level. This overall lookup latency of about 10ms places our solution at the top end of currently available fast DNS resolvers, such as CloudFlare, WordPress.com and others [17].

For service indirection, our platform implementation distributes path information to the ingress and egress points,

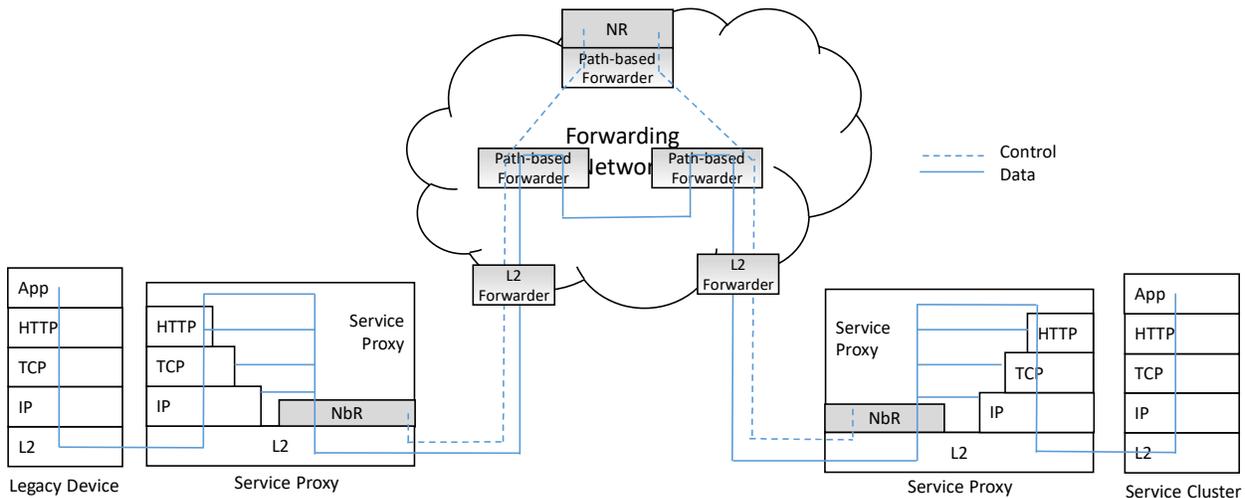

**Fig. 10: Lookup & Indirection Latency Experiment Setup**

allowing for localizing the path computation element. In case of a new service instance becoming available, the update described in Section 3.3.3 is triggered, flagging the name-to-forwarding entry in any ingress and egress that previously discovered the service name as 'stale'. Upon arrival of a new transaction to said service name, the localized path computation does not incur any network latency. With that, service indirection was measured at less than 1ms before the transaction can now be sent to the new service instance.

If the localization feature is switched off, e.g., for not wanting to distribute topology information to the endpoints, the path computation can be triggered upon arrival of the update notification for the service instance. The indirection will take as long as the initial discovery now for cases in which a new transaction arrives at about the same time as the update notification from the NR. In cases, however, where the path computation can take place beforehand, this indirection latency is now reduced down to zero since the new computed path identifier will be found readily available in the local name-path cache of the sending device.

# 5    Deployment Insights

We present in the following two sub-sections insights from realizing and deploying our solution. We firstly outline an initial device implementation, realized on Android terminals, before presenting first trial-based deployments in real world scenarios.

## 5.1   Device Implementations

For the implementation of the flat stack device in Figure1, we developed two choices, shown in Figure 11. The choice on the left-hand side uses the service proxy, usually deployed in proxy devices in support for legacy devices. Any IP-based application is supported, similar to a legacy device in Figure 1, by capturing IP packets through a lookback UDP-based VPN service, terminating the packets in the local service proxy, which in turn forwards the appropriately mapped IP service transactions onto the NbR layer. We utilize the existing IP protocol stack to transfer the named service transaction over a WiFi access point to the L2 forwarder, which we extended to support link-local IP for this purpose. The reason for this IP-based transmission is to offer this implementation choice at the application level, therefore not requiring root access to the device, as would be required when sending Ethernet packets as proposed in Section 3.2. This allows for simple installation as an application, offered through the mobile application store. This choice is directed to support end user trials where test users can bring their own devices and only require installation of our implementation as an application.

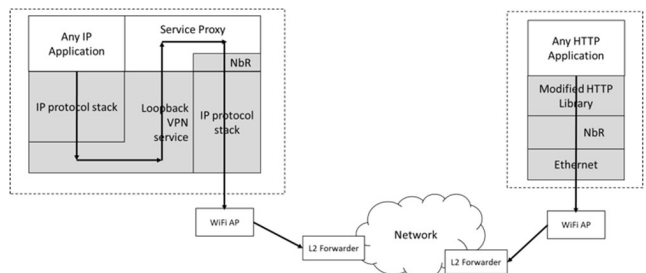

**Fig. 11: Device Implementations**

The right-hand choice in Figure 11 shows our native OS level implementation. Here, the HTTP library has been modified towards the interactions with the NbR layer, following the steps described in Section 3.3. Due to its native level realization, the implementation fully supports the Ethernet-level communication with the L2 forwarder as proposed in Section 3.2. Therefore, the end-to-end communication with other devices occurs without the need for any loopback allowing full insights into possible performance benefits in future systems. The drawback of this design is the needed adaption of low-level libraries, requiring rooting the device itself. Also, HTTP applications must be linked against the modified HTTP library to make use of the named transactions.

We realized both design choices in Figure 11 on Android devices, tested on API levels 24 onwards.

## 5.2   Trial Deployments

Early versions of the network architecture in Figure 1 have been trialed in production networks, such as presented in [7], and in a small city trial with a location-based gaming scenario [8]. Throughout 2019, about 10 further user-facing trials are planned to be conducted in European city deployments with results expected for publication in late 2019. The use cases include location-based gaming, user-generated reporting of events, video streaming on the go and a number of others.

As a specific example of a use case with significant benefits from the multicast capabilities evaluated in Section 4.1, a VR-based tourist guide is being realized over our solution. Here, several users assembled within close proximity consume a virtual reality stream. The playout of said stream is controlled through a master device, which represents the 'tourist guide' in real life guided tours. Hence, upon selecting a specific part of the guide storyline, all users are being directed to start watching at the specific time where information of said storyline is being shown. This creates a naturally strong synchronization between all user devices, transferring the VR stream via HTTP from a central playout server. Due to this synchronization, we expect a significant multicast gain, including over the (WiFi-based) radio link. Such gain is important for the use case provider since current realizations of this use cases (as a commercial product) resort to downloading the VR content to each individual end user devices before the start of the tour, which leads to a significant delay in starting each individual tour, while also incurring costs on the use case provider for the download that is linearly dependent on the number of users per tour. As a result of using our solution, the use case provider not only expects to switch to a runtime playout of the video (therefore lowering said initial delay in starting a tour) but also reducing the costs for transmitting the video due to the constant delivery costs compared to the linearly increasing costs incurred in current systems. Based on the trial insights, expected for mid of 2019, we plan on providing detailed evaluation insights into the specific trial in our future work.

## 6 Related Work

Name-based routing has been investigated for a considerable amount of time in the area of information-centric networking (ICN) [18][19] as a possible replacement of IP routing at Internet scale. The ongoing work in [20] classifies different ICN deployment configuration. The work presented in this paper falls under the 'ICN underlay' deployment classification, utilizing ICN routing capabilities for the provisioning of Internet protocol based services, with migration categories such as edge networks and 'ICN in a slice' being targeted for migration areas. While the solutions here are based on ICN variants developed in [19], work such as that in [21] proposed solutions based on content-centric networking (CCN [19]) albeit with limited use in CDN islands only. There exists a plethora of work on service routing on higher layers, such as Cisco's service routing approach [22], while newer work on service meshes suggest a similar 'dedicated low-latency infrastructure layer' albeit often relying on dedicated proxy instances for each service while our work pursued the integration of such dedicated network service into the end user device.

## 7 Conclusions

Latency and bandwidth utilization pose significant challenges for envisioned new services, particularly in the virtual and augmented reality space. In this paper, we approached this challenge by proposing a backward-compatible name-based routing solution for the edge of the Internet with accompanying mobile device realization. This solution flattens the protocol stacks in end devices, therefore reducing latency, while allowing for stemming the bandwidth costs for scenarios in which concurrent viewing of content allows for multicast delivery instead of the usual, costly, unicast delivery.

We provided not only details of the operations needed to translate existing IP-based protocols onto the proposed name-based routing, but also evaluated our solution along clearly identified opportunities to improve on the key performance indicators of latency and bandwidth. Our upcoming trial insights will ground those evaluation results towards a better understanding on how such novel device and network architectures can lead to significant quality of service and experience improvements. Such understanding is crucially important for our exploitation work in standardization bodies, specifically the Internet Engineering Task Force (IETF) but also the 3rd Generation Partnership Project (3GPP). The former addresses the specific Internet protocol aspects of our solution, while the latter targets the specific inclusion in specification for the currently developed 5th generation of mobile networks, such as the LAN-based forwarding in cellular sub-systems.

Our future work will not only be focused to drive our solution into those standard bodies but also to deepen the development of key elements, most notably the flow and error control aspects of our transport protocol in Section 3.4. For the latter, we specifically focus on solutions optimized for the possibly ad-hoc nature of relationships in multicast use cases, investigating network coding as a possible key technology for its realization.